# Nepal Himalaya Offers Considerable Potential for Pumped Storage Hydropower

Rupesh Baniya[1], Rocky Talchabhadel[2*], Jeeban Panthi[3], Ganesh R Ghimire[4], Sanjib Sharma[5], Prithvi Dhwoj Khadka[6], Sanghoon Shin[7], Yadu Pokhrel[8], Utsav Bhattarai[9], Rajaram Prajapati[10], Bhesh Raj Thapa[11], and Ramesh Kumar Maskey[12]

[1]Institute of Engineering, Pulchowk Campus, Lalitpur, Nepal; *rupesh.baniya480@gmail.com*

[2]Texas A&M AgriLife Research, Texas A&M University, El Paso, TX, USA; *rocky.talchabhadel@ag.tamu.edu*

[3]Department of Geosciences, University of Rhode Island, Kingston, RI, USA; *jeeban_panthi@uri.edu*

[4]Environmental Sciences Division, Oak Ridge National Laboratory, Oak Ridge, TN, USA; *ganeshghimire1986@gmail.com*

[5]Earth and Environmental Systems Institute, The Pennsylvania State University, University Park, PA, USA; *svs6308@psu.edu*

[6]Department of Hydrosciences, Technische Universität Dresden, Tharandt, Germany; *prithvidhwoj.khadka@gmail.com*

[7]Earth System Science Interdisciplinary Center, University of Maryland, College Park, MD, USA; *shinsa11@umd.edu*

[8]Department of Civil and Environmental Engineering, Michigan State University, East Lansing, MI, USA; *ypokhrel@egr.msu.edu*

[9]Institute for Life Sciences and the Environment, University of Southern Queensland, Toowoomba, Queensland 4350, Australia; *Utsav.Bhattarai@usq.edu.au*

[10]Department of Geology and Environmental Science, University of Pittsburgh, PA, USA; *rajaram@smartphones4water.org*

[11]Universal Engineering and Science College, Pokhara University, Lalitpur, Nepal; *bthapa.ioe@gmail.com*

[12]Nepal Academy of Science and Technology, Nepal; *drrameshma@gmail.com*

*Corresponding author: Rocky Talchabhadel (*rocky.talchabhadel@ag.tamu.edu*)

**Abstract**

There is a pressing need for a transition from fossil fuel to renewable energy to meet the increasing energy demands and reduce greenhouse gas (GHG) emissions. The Himalayas possess substantial renewable energy potential that can be harnessed through hydropower projects due to its peculiar topographic characteristics and abundant water resources. However, the current exploitation rate is low owing to the predominance of run-of-river hydropower systems to support the power system. The utility-scale storage facility is crucial in the load scenario of an integrated power system to manage diurnal variation, peak demand, and penetration of intermittent energy sources. In this study, we first identify the potential of pumped storage hydropower across Nepal (a central Himalayan country) under multiple configurations by pairing lakes, hydropower projects, rivers, and available flat terrains. We then identify technically feasible pairs from those of potential locations. Infrastructural, environmental, operational, and other technical constraints govern the choice of feasible locations. We find the flat land-to-river configuration most promising than other configurations. Our results provide insight into the potential of pumped storage hydropower and are of practical importance in planning sustainable power systems in the Himalayas and beyond.

# 1. Introduction

The global energy sector is the largest contributor to greenhouse gas (GHG) emissions holding the key to averting the impacts of climate change [1]. The 26th United Nations Climate Change Conference of the Parties (COP26) recommended necessary actions to limit the global rise in average temperature below 2 °C from pre-industrial times and to pursue efforts to restrict it to 1.5 °C [2,3]. The shift toward Net Zero Emissions by 2050 requires nations to unite and efficiently implement energy and climate change policies, including a massive transformation of the energy sector [4,5]. Increasing the deployment of renewable energy sources is crucial for this transformation [6–8]. Countries with fossil fuel as a primary energy source have a crucial role in significantly mitigating GHG emissions by switching to renewables [3]. Hydropower, the world's largest source of renewable energy, holds the key to this transformation [2,5], given the variable nature of other renewables such as wind and solar. Hydropower is one of the clean, most cost-effective, and most flexible energy storage technology that can help to ensure a reliable and secure energy supply [9]. The assessment led by the International Energy Agency (IEA) and the International Renewable Energy Agency (IRENA) estimates that at least 850 GW of hydropower must be produced to keep global warming below 2°C. The figure needs to be doubled to meet the Net Zero emissions target (i.e., below 1.5°C) by 2050 [5].

With the rapidly evolving electric grid system due to the influx of wind and solar, there is a need for large-scale energy storage [10,11]. For the global electricity market, hydropower is the least expensive and most efficient large-scale energy storage alternative compared to other technologies such as batteries, hydrogen, and flywheel [8,12–14]. Pumped storage hydropower (PSH; **Fig. 1**) functions like a giant battery allowing the much-needed reliability and flexibility in the electric grid system [10]. This helps to reduce the need for fossil fuel-based energy sources, which is critical for meeting a single SDG [15], for example, SDG 7 (affordable and clean energy) and SDG 13 (climate action). Additionally, PSH can provide additional jobs and economic benefits to local communities, thus contributing to SDG 8 (decent work and economic growth), and meeting SDG2 (Zero Hunger) [16]. Finally, it can also help to protect local ecosystems by mitigating the impacts of extreme weather events, thus contributing to SDG 15 (life on land). Therefore, the energy system is ultimately relevant to multiple SDGs [17]. PSH alone accounts for ~90% of the world's grid-scale storage applications (160 GW) [5]. Importantly, PSH's ability to store large-scale off-peak, excess, or unusable electrical energy and to facilitate optimal production and consumption with grid stabilization [18,19] makes it the most adopted energy storage technology. PSH is crucial for sustainable transformation in energy due to its ability to balance electricity supply and demand, as well as its potential to store large amounts of energy [8]. It is an important tool in the transition towards a low-carbon energy system, as it can help to reduce the need for fossil-fuel-based electricity generation and provide a way to store excess electricity generated from renewable energy sources [20]. Hence, it is critical to assess feasible locations for such storage projects. Global assessment of the off-PSH identified 616,000 promising locations with a combined storage potential of 23 million GWh [21]. Several regional assessments have shown similar prospects for PSH in different parts of the world, including Turkey [22], the United States [23], France [24], and Iran [25], among others.

The potential site mapping for PSH involves identifying suitable pairs of lower and upper reservoirs followed by an estimation of electricity storage capacity [8,23,24,26]. Globally, various approaches have been proposed for identifying the appropriate locations for PSH projects. Ahmadi and Shamsai [19] and Jiménez Capilla et al. [20] determined the optimal location of an upper reservoir in the proximity of a reference hydropower reservoir using

Multiple-Criteria Decision-Making (MCDM) techniques. Such techniques have also been used to evaluate the best alternative from predetermined PSH locations [29–31]. Lu and Wang [32] investigated narrow valleys instead of flat areas in Tibet for possible use as reservoirs in integration with existing lakes. For off-river pumped hydro schemes, Lu et al. [25] developed an advanced Geographic Information System (GIS) algorithm to identify two reservoir models (i.e., dry-gully and turkey's nest) in South Australia. Using the GIS-based model and topographic information, Gimeno-Gutiérrez and Lacal Arántegui [34] demonstrated significant theoretical potential for PSH in several European countries. Furthermore, recent studies explored the utilization of natural depressions [24], seawater [35], and wastewater treatment facilities [36] as potential configurations for PSH reservoirs. Most of these studies concentrated on examining potential under topological relations between reservoirs reported by the Joint Research Center of the European Commission [37]. The utilization of rivers as an upper/lower reservoir in the PSH scheme is less explored. Considering the topography of Iran, Ghorbani et al. [25] divided the river into a set of points at 40km intervals (site for the lower reservoir) and searched for suitable flatland for an upper reservoir. Görtz et al. [38] developed a new algorithm to locate suitable ring dam locations along rivers and shorelines. However, this method explores less in terms of potential connection between upper and lower reservoir points and lacks the quantification of energy storage capacity. Our study employs a point based search along river network for potential utilization of river (site for lower/upper reservoir) in the mountainous region. This approach is capable of estimating pumped energy storage capacity of rivers in combination with the nearby lakes and flatlands.

The Nepal Himalayas possess an abundance of renewable energy potential, primarily through hydropower [39,40]. Hydropower energy's contribution to the electric grid in the region is predominantly from the run-of-river (RoR) hydropower plants [41]. Numerous previous studies have examined RoR and storage-type hydropower projects in Nepal [42–45]. Moreover, to complement a large number of existing and planned ROR hydropower plants [46,47], PSH could be an efficient and cost-effective energy storage alternative[48]. Diverse topographic conditions, sharp elevation gradient, high stream power, and perennial water source facilitate a huge potential for hydropower development in the central Himalayan region [49]. A few studies (e.g., [48,50,51]) exist on the potential of PSH in the Nepal Himalayas, but much fewer than the traditional RoR hydropower schemes [52–56]. Nepal Himalayas provide an ideal testbed to study pumped storage systems given high topographic gradients, large flow fluctuations, and prevalent energy demand patterns.

The Global Pumped Hydro Storage Atlas [57] used GIS-based algorithms [33] to identify around 2,800 potential locations in the Himalayan country Nepal for off-river schemes, such as two reservoirs located in proximity but at different altitudes and connected by a pipe or tunnel [21,48]. Recently, there have been some initiatives to explore PSH in the Himalayan region. For example, the Government of Nepal is currently exploring several possible locations for PSH, such as Begnas-Rupa (150 MW), Lower Seti (104 MW), and Kulekhani (100 MW) [58]. Previous studies [48,50,51] provide important insights into the potential PSH locations in Nepal, but they are focused only on exploring limited configurations at a location such as lake-to-lake connections [51] and pump-back systems between hydro-project reservoirs [58]. Therefore, a nationwide identification of potential locations for PSH considering a wide range of configurations (e.g., lakes, hydropower projects, rivers, and available flat terrain) is crucial to developing reliable decision support systems for the sustainable utilization of water resources.

This study is the first of its kind to explore the suitability of pumped storage schemes and their potential in the Himalayas. Based on the diverse topographic characteristics of the

Himalayan region, this study considered the full utilization of natural lakes, flatlands, and rivers employing several reservoir configurations for developing PSH. We employ a geospatial model to identify the viability of PSH. The major objective of this study is to characterize the baseline energy potential of PSH across the Himalayas. Here, we address three key research questions: (i) What are the theoretical, technical, and exploitable potential of PSH in the Himalayas? (ii) What is the preferred reservoir configuration in the Himalayan topography?, and (iii) How do topography and hydroclimatic conditions affect the spatial distribution of PSH? This study provides an entry point for discussion among energy planners, decision-makers, and modelers to develop sustainable energy systems. The proposed approach could also be employed in other global regions.

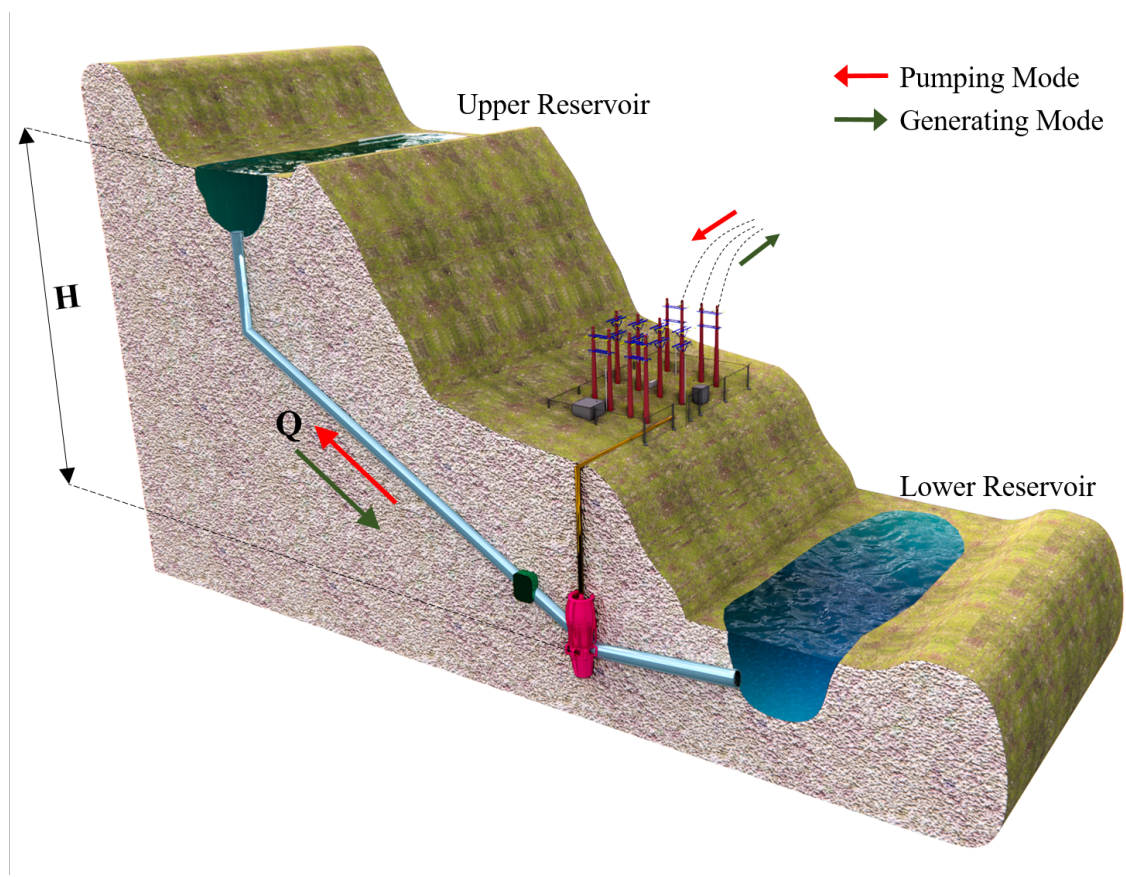

**Fig. 1.** Schematic diagram of a typical pumped storage hydropower system (adapted from ILI Group; https://www.ili-energy.com/why-pump-storage). Pumped storage schemes are designed to operate in two modes: i) pumping water to an upper reservoir using surplus energy (shown by red arrows) and ii) generating electricity by releasing this water back to a lower reservoir during high demand (shown by green arrows). H: Hydraulic Head, Q: Flow

## 2. Materials and Methods

### 2.1. Study Area and Data

In Nepal, there are over 6000 rivers and rivulets with a water storage capacity of over 200 billion m$^3$ [59]. The country is characterized by a complex terrain with many glaciers, rivers, valleys, and lakes. Of 5358 lakes, 2315 are glacial [60]. The high topographic gradient with elevation varying from 60 to 8848 m above sea level (asl) provides both opportunity and challenge in developing transport and grid infrastructure. However, a PSH plant is not feasible at a very high altitude from a construction and operation point of view. Also, prerequisites like road access and transmission infrastructure are challenging to develop. In this study, we excluded regions with elevations greater than 5000 m asl. Such exclusion has been adopted in RoR hydropower potential study by Water and Energy Commission Secretariat, Nepal [40]. Various spatial, hydrography, and infrastructure data used in this study are illustrated in **Table 1**.

**Table 1.** Geospatial data and their sources used in this study

| Dataset | Source |
| --- | --- |
| Digital Elevation Model | SRTM, USGS *https://earthexplorer.usgs.gov* |
| Lakes | OCHA [61] *https://data.humdata.org/dataset/nepal-watercourses-rivers* |
| River Network | Regional Database System, ICIMOD [62] *https://rds.icimod.org/* |
| Road Network | OCHA [63] *https://data.humdata.org/dataset/nepal-road-network* |
| Transmission Network | RPGCL, Nepal [64] |
| Protected Areas | DNWPC, Nepal |

*SRTM*: Shuttle Radar Topography Mission; *USGS*: United States Geological Survey; *ICIMOD*: International Center for Integrated Mountain Development; *RPGCL*: Rastriya Prasaran Grid Company Limited; *DNPWC*: Department of National Parks and Wildlife Conservation.

We assessed hydrometeorological characteristics (precipitation, temperature, and streamflow) of PSH potential locations using average values computed from 40 years of data. We used monthly climate (precipitation and temperature) data from the recent 40 years (1981-2020) of the TerraClimate dataset (spatial resolution ~4 km; Abatzoglou et al. [65]). Streamflow data were taken from Shin et al. [66], spanning 40 years (1979–2018) at ~5 km spatial resolution. We used 10 percentile ($Q_{10p}$: low flow), 50 percentile ($Q_{50p}$: median flow), 90 percentile ($Q_{90p}$: high flow), and average ($Q_{avg}$: mean flow) streamflow data at the PSH potential locations for those configurations that involve rivers (i.e., flat land or lake to the river). Hydrometeorological characteristics of PSH potential locations were categorized for different elevations bands (EBs) above sea level: EB1 (0 to 500 m), EB2 (500 - 1000 m), EB3 (1000 - 2000 m), EB4 (2000 to 3000 m), and EB5 (3000 to 5000 m).

*2.2. Reservoirs Selection*

A minimum of two reservoirs at a certain elevation difference is required for integrating pumping and generating facilities (**Fig. 1**). These reservoirs can be either natural lakes or artificial storage facilities constructed by damming the river or excavating suitable flat land. A PSH scheme can be established by combining these natural and artificial features at varying altitudes. Based on varying topographic settings in the Himalayan region, we proposed four schemes for a combination of these features as shown in **Fig. 2**. We studied storage potential in each scheme that utilizes three types of prospective reservoir locations; natural lake "L", flatland "F", and river "R" by applying the methodological framework shown in **Fig.3**. All the natural and artificial storage facilities were screened with criteria of minimum volume required to achieve the energy storage threshold, discussed later in the following section.

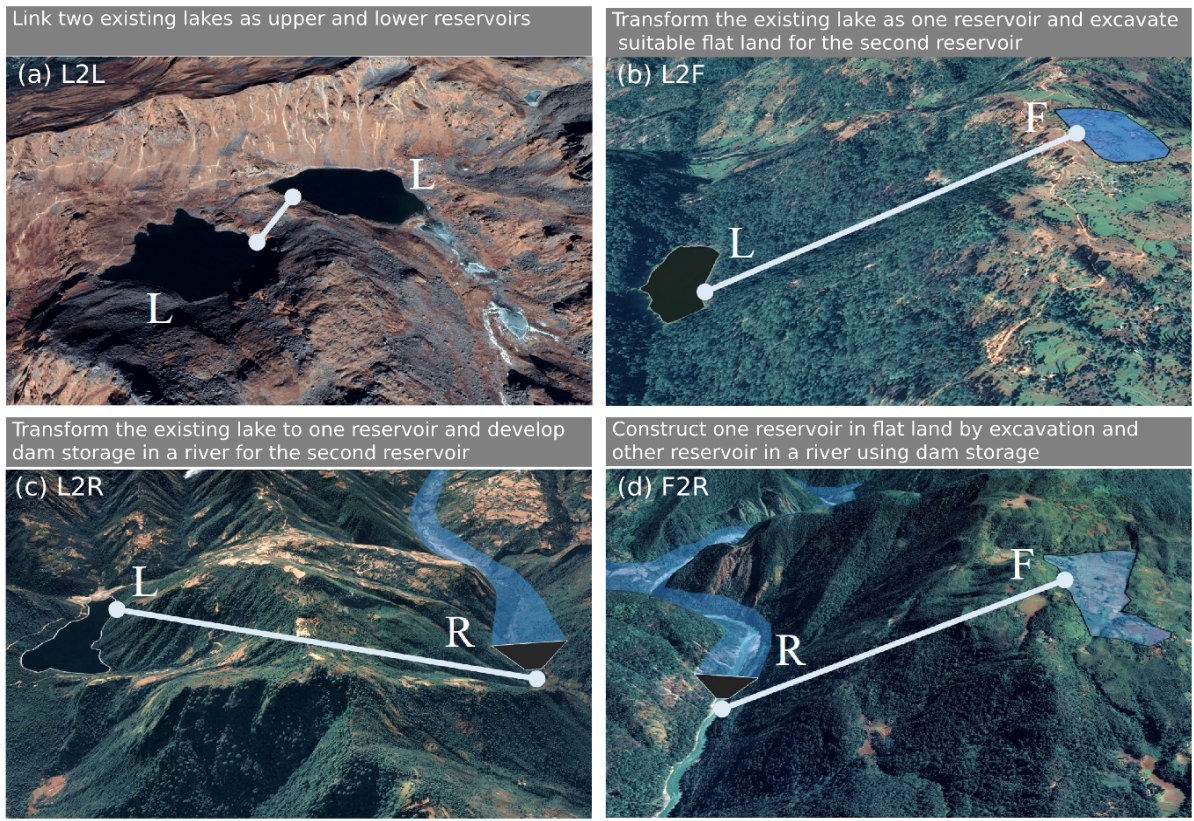

**Fig. 2.** Reservoir configurations investigated in the study. Maps are prepared on the Google Earth platform.

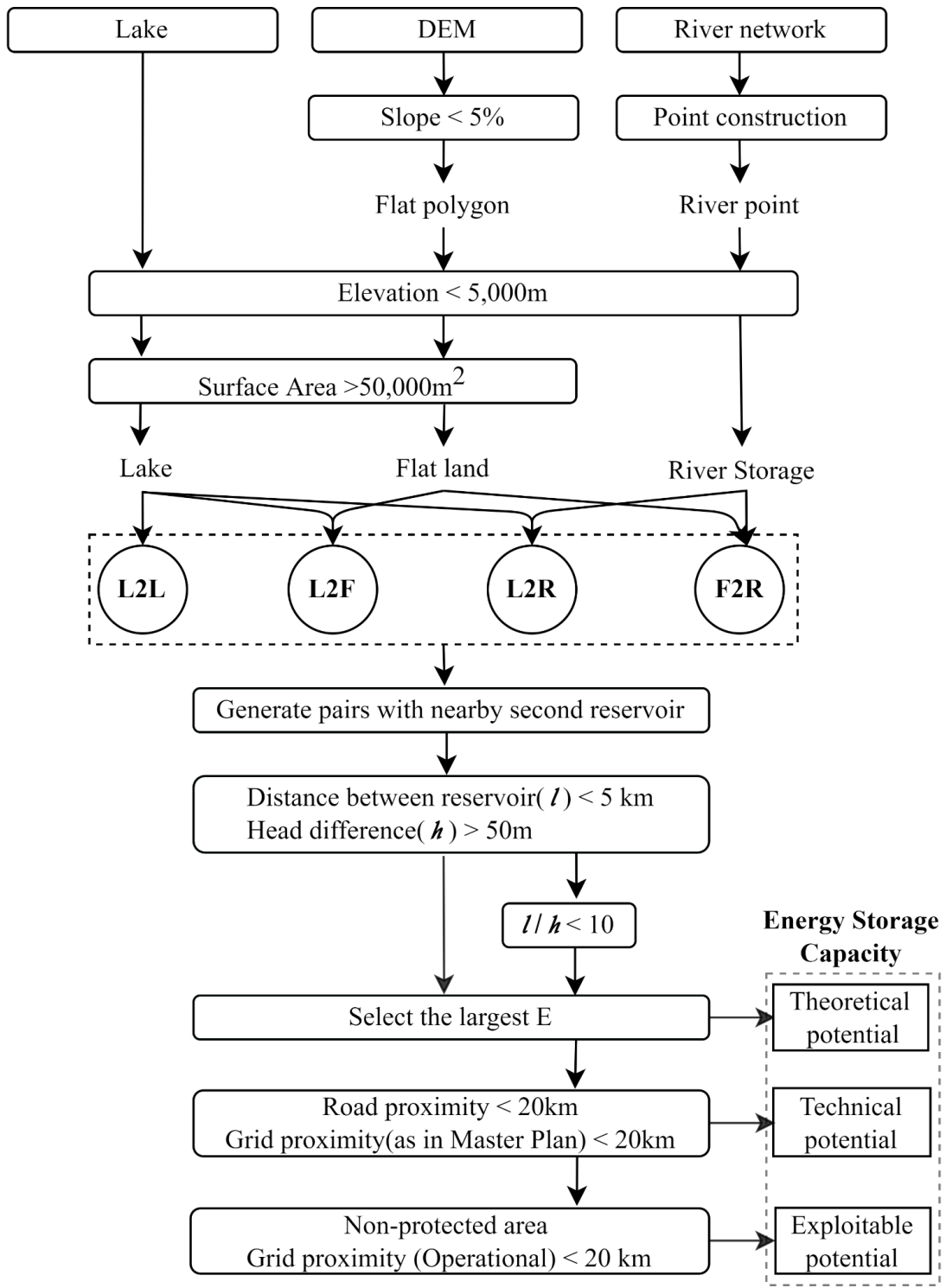

**Fig. 3.** Overall methodological flowchart of this study. Based on topography, four reservoir schemes are assessed: (i) Lake to Lake (L2L), (ii) Lake to Flat land (L2F), (iii) Lake to River (L2R), and (iv) Flat land to River (F2R). Proximity to the road and non-protected areas are used to filter exploitable energy from the theoretical potential locations. Detailed descriptions of these reservoir schemes are in Fig. 2. DEM: Digital Elevation Model, E: Energy Storage Capacity

*2.3. Energy Storage Capacity Estimation*

The hydroelectricity storage potential of a PSH plant is directly proportional to the head difference between upper and lower reservoirs and the utilizable volume of water:

$$E = \eta \rho V g H / (3600 \times 10^9) \quad \ldots\ldots\ldots\ldots\ldots(1)$$

where, E = energy (GWh), ŋ = overall efficiency, $\rho$ = density of water (1000 kg/m$^3$), V = usable volume of an upper reservoir (m$^3$), g = gravitational constant (9.8 m/s$^2$), and H = head difference (m) between upper and lower storage reservoirs.

The performance of the cyclic operation in pumping and generating mode can be better understood by round-trip efficiency. A round-trip efficiency of a typical PSH system ranges between 70 and 80 % [67,68]. The water storage capacity of a reservoir is highly site-specific and dependent on reservoir characteristics, including storage-elevation curve, type, and purpose. Among the two reservoirs in a PSH plant, the reservoir with minimum water storage capacity governs the energy storage potential [32]. We computed the usable volume by a PSH plant as the product of the surface area and utilizable water depth of such a limiting reservoir.

For the L2L and L2F configurations, we used the minimum surface area between the upper and lower reservoirs. However, for a scheme involving a river as one reservoir, we assumed sufficient flow in the river and the second reservoir as a constraint. So, we used the surface area of the second reservoir (either lake or flat land) in the potential calculation of L2R and F2R locations. In the case of a natural lake, consideration of higher utilizable depth creates environmental, social, and technical complications although it may lead to an overestimation of usable water volume. Connolly et al. [69] proposed the construction of a new reservoir of 20 m height on flat land with reference to the existing PSH reservoir like Taum Sauk in Missouri, USA, and Turlough Hill, Ireland. Many authors used the same depth for schemes involving the construction of new reservoirs in flat areas. In mountainous terrains such as Nepal, the development of such massive reservoirs on flat land may not be geologically and economically favorable. Further, due to high vulnerability to seismic activity, such depth may lead to Reservoir Induced Seismicity (RIS) [70,71]. Therefore, 2 m of the depth of the existing lake and new reservoir in excavated flat land was taken as utilizable depth for all schemes.

We aimed at developing a model that can detect even the small PSH site with an energy threshold of 10 MWh (a 1 MW plant active for 10 hours). In all schemes, a minimum head difference between the upper and lower reservoir was set to be 50m. In this connection, for a 2 m utilizable depth, minimum storage of 100,000 m$^3$ is required. So, our study is limited to only those reservoirs with a surface area greater than 50,000 m$^2$. The minimum volume selection is consistent with the constraint adapted by Gimeno-Gutierrez and Lacal-Arantegui (2013).

*2.4. Definition of Energy Storage Potential*

Theoretical potential corresponds to all the locations that satisfy the fundamental requirement/provision, including head difference and water storage capacity. Theoretical potential captures the energy storage capacity of all the locations; each one with two reservoirs at different altitudes and certain water volumes that can be used in the cyclic operation. These locations are assumed to be harnessed to their full potential, i.e. operate at 100% efficiency.

The technical potential is deduced considering topographical, operational, and infrastructure constraints. The topographical characteristics of reservoirs in PSH plants can be measured by the ratio of the distance between them and the head difference, denoted as l/h.

The lesser the l/h ratio, the more economically attractive would be the site [19]. Locations with l/h lesser than 10 are considered technically more suitable [37,72]. To account for mechanical energy losses, an efficiency of 80% was assumed in calculating technical energy storage potential. Infrastructure facilities like transportation are required for construction and operation. The powerhouse in the PSH site should be connected to the grid to transmit electricity. Locations that are in less than 20 km proximity to road and grid facilities are considered to satisfy the infrastructure constraint. We identified technical potential locations in compliance with the transmission facility as envisaged by the Transmission Network Master Plan of Nepal [55].

Exploitable potential represents the technical locations that can be exploited/realistically developed now with the existing grid facility and are free from environmental restrictions. The Working Policy on Construction and Operation of Development Projects in Protected Areas [73] guides the development of river-based hydro blocking/diverting water, but with many restrictions. In the present scenario, we assumed the development of a PSH plant would obstruct the objective of such areas. So, the technical locations located in such areas are excluded to find out the exploitable potential.

2.5. Reservoir Configuration

We configured an integrated modeling system (**Fig. 3**) to assess the potential of different reservoir schemes (**Fig. 2**). The modeling framework undergoes scheme-specific input processing, reservoir pairing, and constraint application. We chose an appropriate reservoir, paired with a second reservoir (either upper or lower), and computed the energy storage capacity of that particular combination. Obtained locations were sequentially tested under user-defined constraints for their theoretical, technical, and exploitable viability.

*2.5.1. Prospective Reservoir*

Databases of reservoir facilities are required as input for the model. Except for lakes, other reservoir databases are created by the geospatial operation of the Digital Elevation Model (DEM) and river network. For flat land, polygons with less than 5 % slope are obtained from DEM. Area and the average elevation of the lake and flatland are then computed. By filtering the reservoir features with a surface area greater than 50,000 m² located below 5000 m asl elevation, prospective reservoirs are obtained. For schemes utilizing rivers, points are constructed at an interval of 1 km along the river network. The elevation of these river points is extracted from the DEM. These points are considered to be the on-river storage site where the storage facility can be created by dam construction. This dam can also be used for operating a conventional RoR hydropower plant in parallel with the PSH scheme.

*2.5.2. Selection of Reservoir Pairs*

Each prospective reservoir should be paired with the second reservoir for a PSH plant. The first letter in scheme notation denotes the prospective reservoir, while the second letter denotes the second reservoir in a pairing. For example, in the F2R scheme, the model searches numerous river locations to pair with each flat land. For each prospective reservoir, all the nearby reservoirs within a 10 km search radius (either at lower or upper elevation) were found. This provides multiple options for prospective reservoirs to pair with a second reservoir to form a PSH site. Pairs satisfying 50 m minimum head and 5 km maximum distance criteria were selected. The pair offering the largest energy storage capacity was selected for theoretical potential calculation. However, for the technical potential study, we selected pairs with an l/h ratio of less than 10. And only then, the pair with the largest storage capacity was chosen as a PSH site configuration for further technical analysis.

Different criteria were applied in the model to deduce storage potential under the technically feasible category. Access to infrastructural facilities like transportation and the grid was examined in the model. We adopted the transmission network master plan of the Government of Nepal for technical potential calculation [64]. A buffer of a 20 km radius was created around the major road network and substations. Locations located inside the intersection of two buffer zones are technical locations. Then, technical locations located inside protected areas were excluded for evaluating the exploitable potential. The remaining locations inside a 20km buffer zone around existing substations (operational) give an insight into exploitable potential.

## 3. Results

### 3.1. Storage Potential

Utilizing the two existing lakes (L2L scheme), we observed a total of 89 potential locations, with a combined theoretical storage capacity of 11.3 GWh. However, all theoretical locations were of capacities less than 1 GWh. Technically, 29 locations were found to be suitable, with a potential of about 4.1 GWh. We observed that the most technically feasible locations (greater than 0.1 GWh, shown in green squares in **Fig. 4**) were located in the northeast region of the country. Only one exploitable site was found with a larger storage capacity, i.e., 0.3 GWh (between Begnas and Rupa Lakes in Northeast Nepal). This project is currently under study by Nepal Electricity Authority (NEA 2020).

The theoretical storage capacity under the scheme of pairing lakes with flat land (L2F) was found to be about 7.9 GWh from a total of 37 locations. Two locations were technically viable, with a cumulative capacity of 0.9 GWh. Only two theoretical locations had a larger storage capacity greater than 1 GWh. However, none satisfied technical constraints. The exploitable potential was found to be about 0.2 GWh, from only two locations. Theoretically, 205 lakes could be connected with rivers by incorporating PSH infrastructure to store 276.5 GWh of energy. The technical potential was estimated to be about 65.1 GWh, from a total of 88 locations. The majority of technical locations (about 80%) indicate storage capacities between 0.1 and 1.0 GWh. A total of six locations could be counted as the exploitable category, with a cumulative storage capacity of 6.4 GWh. For pairing a lake with a second reservoir, river points are a readily available option to the option of flat land and another lake. Therefore, the L2R configuration shows greater potential than those utilizing lakes.

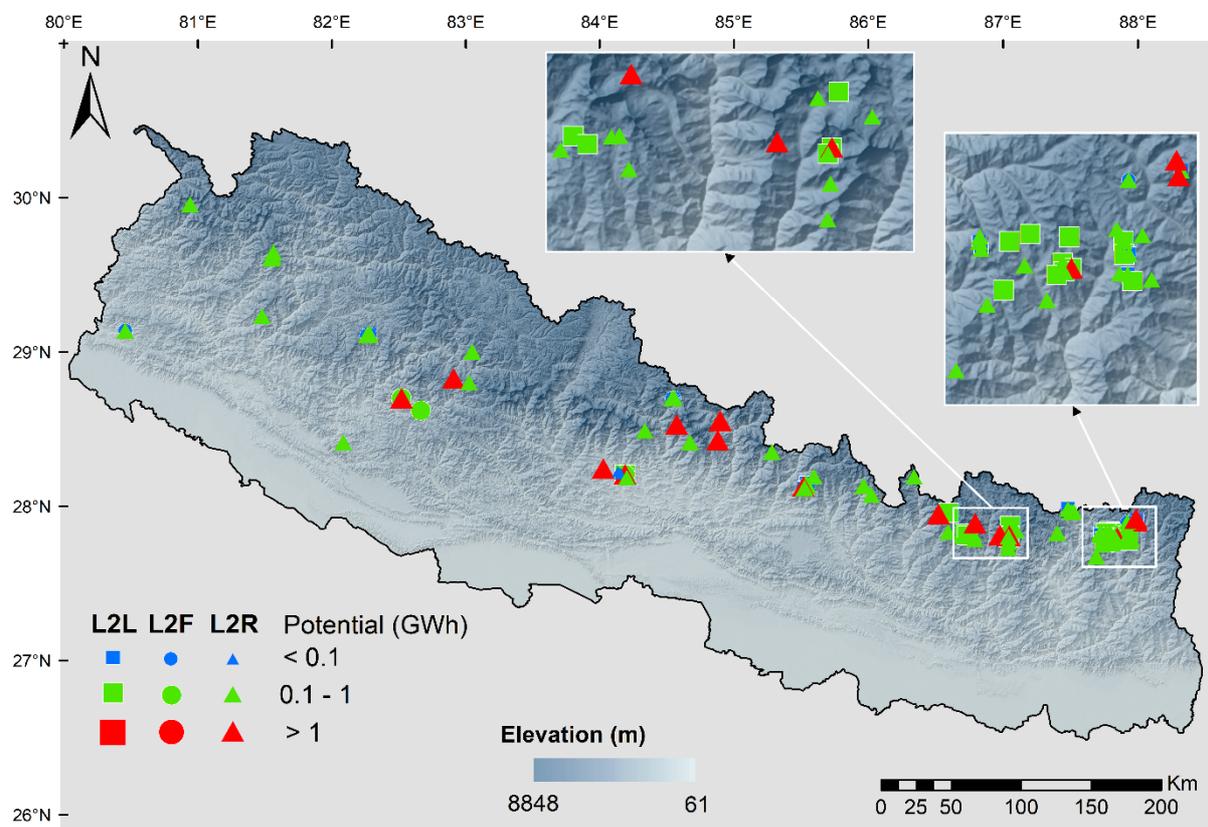

**Fig. 4**. Spatial distribution of technically feasible Lake to Lake (L2L shown in squares), Lake to Flat land (L2F shown in circles), and Lake to River (L2R shown in triangles) locations across Nepal. The shaded background represents the underlying topography.

We identified 7,440 potential flat land locations under the F2R configuration. Among them, 6134 locations were found theoretically viable, yielding a total energy storage capacity of 2716 GWh. **Fig. 5** shows 1739 technically potential locations that could provide an estimated energy storage capacity of 1198.8 GWh. And, out of 1739, 1184 locations could be exploited, with a storage capacity of 897.9 GWh. Noticeably, the exploitable F2R locations were substantially larger and more widely distributed across the country compared to other configurations (**Fig. 5**).

**Fig. 5** shows that 68% of the technically feasible PSH locations (n = 1188) have the potential to provide energy storage between 0.1 and 1 GWh. We observed that these locations were mostly distributed between mid-hills and southern plains. Because of the relatively greater availability of the flat lands, the larger potential locations with energy storage capacity > 1 GWh (n = 174, i.e., 10%) were mostly identified in the southern plains.

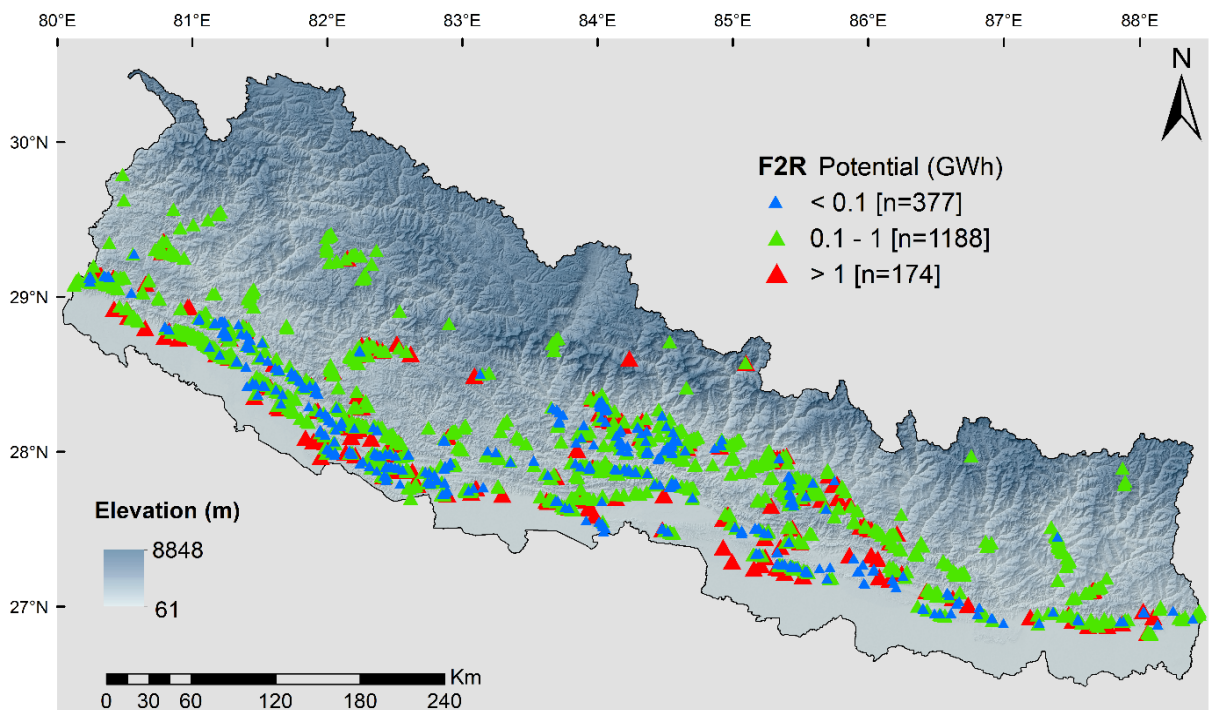

**Fig. 5.** Spatial distribution of technically viable Flat land to River (F2R) locations across Nepal. The shaded background shows the topography.

*3.2. Characterization of Potential Locations with Hydroclimate and Topography*

**Fig. 6** shows the distribution of hydroclimatic characteristics of technically potential PSH locations. Most F2R schemes were located in lower regions, i.e., EB1 (N=1063) and EB2 (N=491). The average streamflows ($Q_{avg}$) at these locations were 173 m$^3$/s and 84 m$^3$/s, with a high standard deviation of 302 m$^3$/s and 121 m$^3$/s, in EB1 and EB2, respectively. The mean annual precipitation at these locations was approximately 1670 mm. Since these locations are at low altitudes, the annual average temperature is over 20°C and the climate is mostly tropical. Detail for each scheme at different elevation bands is also provided in **Fig. 6**. Out of 88 technically potential L2R schemes, 78 locations were located in EB5. There were no locations in EB1, six in EB2, three in EB3, and one in EB4. These regions also had limited precipitation, with mean annual precipitation below 850 mm. For L2F and L2L, we showed the distribution of precipitation and temperature at the technically viable locations. All technically potential L2L locations were located in EB5 (N=28), except one in EB2. Likewise, L2R locations in EB5, these locations had less precipitation and freezing temperatures. Only nine locations were found suitable for technically viable L2F schemes. Out of nine, five are in EB5 (mean annual precipitation less than 650 mm and an average freezing temperature), two in EB2, and two in EB3.

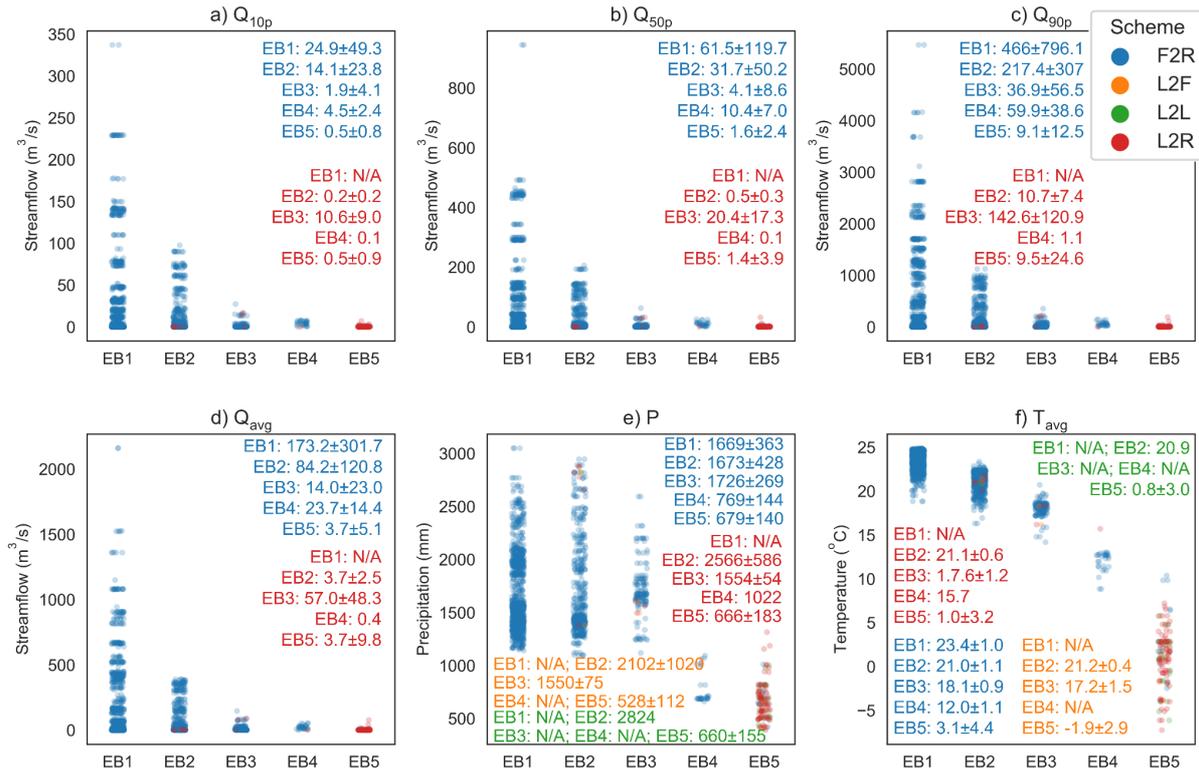

**Fig. 6.** Strip distribution of technically viable pumped storage hydropower (PSH) schemes at different elevation bands (EB1: 0 to 500 m, EB2: 500 to 1000 m, EB3: 1000 to 2000 m, EB4: 2000 to 3000 m, and EB5: 3000 to 5000 m above sea level) across Nepal. The four PSH schemes are Flat land to River (F2R) in blue, Lake to Flat land (L2F) in orange, Lake to Lake (L2L) in green, and Lake to River (L2R) in red. Detailed descriptions of these PSH schemes are in **Fig. 2**. Text inside each plot represents the mean (u) and standard deviation (σ), represented as u±σ, and corresponding texts are depicted in their respective colors. The streamflows (i.e., $Q_{10p}$, $Q_{50p}$, $Q_{90p}$, and $Q_{avg}$) for L2L and L2F are not assessed since these schemes do not involve river streamflow.

## 4. Discussion

A PSH facility can be established in many ways, for example, by installing a pump-back system between multiple reservoirs or using flat land as a second reservoir in the vicinity. The current study presented an integrated modeling framework to examine theoretically and technically viable PSH locations. The model can further explain economic potential by incorporating cost and benefit analysis parameters.

Furthermore, the exploitable potential evolves with time and can depend on various factors, including legal frameworks, infrastructures, and other site-specific conditions including topography. For example, our analysis resulted in 1,193 exploitable PSH locations yielding 904.8 GWh in generation mode with an annual energy potential of 330,252 GWh. This estimate is about a six-fold increase from the estimate (i.e., 58000 GWh) by WECS [40] that used 35 dam locations in the river for conventional hydropower reservoir projects highlighting the importance of a comprehensive framework for such assessments. However, the study region still lacks a market framework, including peak-load pricing. As the monetary benefit from PSH projects predominantly depends on the price difference between pumping and generating energy [74], a proper regulatory framework needs to be designed. In addition, the availability of grid infrastructure is important. The grid infrastructure seems to rapidly expand to meet the increasing energy demands. We found that numerous potential locations are located inside protected areas. And, as the prevailing acts, legal frameworks, and policies do not allow large-scale construction [73], these locations would not result in energy generation.

Our results show that the PSH locations are mostly around mid-hills across the country. Global Pumped Hydro Storage Atlas (GPHSA; [57]) showed a lower concentration around the central and eastern regions' mid-hills and a higher concentration around the mid-hills of the western region. These differences could be partly due to additional configurations of lakes, flatlands, and rivers explored in our approach than GPHSA. The GPHSA explored off-river configurations using an upper reservoir in high hilly areas rather than in a river valley. In contrast, we explored additional configurations with upstream reservoirs using lakes or flat lands (i.e., L2L, L2F, L2R, and F2R). Also, GPHSA [57] indicated additional locations in northern high altitudes. These higher altitude locations were limited in our approach because of the elevational constraints employed in our modeling approach. As these locations are situated over 5000 m asl, technically and economically, it would be quite difficult to construct the necessary infrastructures. Furthermore, due to the high altitude, L2R schemes in EB5 had a mean annual average temperature slightly above or around freezing temperature, indicating chances of freezing lakes during the winter season and limiting the dry season to produce energy effectively. However, due to a lower temperature, evaporation losses during the summer would be less across the lakes, which is beneficial for hydropower production. Himalayan rivers located at these high locations have smaller streamflow. For example, the high flows (i.e., $Q_{90}$) for L2R schemes were only 10 m$^3$/s. In the case of F2R, larger flat areas indicate the larger energy storage potential in the southern plains of Nepal, by constructing reservoirs. Such wide distributions of high-potential PSH locations provide opportunities to add flexibility to the grid system at both local and national scales.

Technically, PSH is a unique kind of hydropower project based on the water cycle between two reservoirs. Once the water is stored, the same water is used for generating and pumping. Integrated planning of PSH reservoirs will enhance the overall ecosystem. Water conservation in such areas can be a tool to minimize the impact of climate change as well. Through new legal approaches and proper guidelines to address environmental aspects, specific PSH locations can be developed after a detailed study. Regulating agencies like Nepal Electricity Authority can develop and operate PSH projects as a daily load-balancing tool. It facilitates the optimized operation of its RoR and storage projects. Also, it helps to manage anticipated spilled energy in the near future.

## 5. Conclusion

In this study, we configured a geospatial model to identify the potential of PSH across the Nepal Himalayas under multiple configurations by pairing lakes, hydropower projects, rivers, and available flat terrain, and consequently estimate the energy storage capacity. Our study applied a novel approach of reservoir pairing for each prospective reservoir to form the technically suitable pair from multiple configurations of the second reservoir. Applying technical constraints, we obtained technically feasible locations with grid access based on the Government of Nepal's master plan of the transmission network. Finally, the exploitable locations and current energy storage potential were identified by employing environmental constraints and existing grid facilities. We summarize below the key findings from this study:

- The exploitable F2R locations were substantially larger and more widely distributed across the country compared to other configurations.
- The overall distribution of technically and theoretically feasible locations is more concentrated in mid-hills and southern plains. High-altitude mountain regions characterized by low precipitation, average temperature close to freezing point, and complex topographic features demonstrate relatively limited potential for PSH.
- In total, 3012 GWh is estimated as theoretical potential and 1269 GWh (42% of theoretical) as technical potential across the Nepal Himalayas.

PSH's large potential for energy storage in the Nepal Himalayas is a precursor for Nepal to become a seasonal power hub in the region. Furthermore, in the South Asia region, there is a seasonal complementarity in the power system among the countries [75]. Despite implementation at the national scale, the methods and models developed in this study are quick, simple, and generalizable, making their application feasible at regional and global scales. Note that the identified PSH potential might alter with future environmental and anthropogenic activities such as hydroclimatic variability, land use land cover changes, and new infrastructure developments (e.g., dams and reservoirs). Developing PSH infrastructure often requires a higher upfront investment. More research is needed on the integrated PSH system, grid connections, and economic model to inform the future development of PSH.

**Authorship Contribution Statement**

**Rupesh Baniya**: Conceptualization, Methodology, Software, Formal Analysis, Investigation, Writing- original draft, Visualization. **Rocky Talchabhadel**: Conceptualization, Methodology, Formal Analysis, Writing – review & editing, Visualization. **Jeeban Panthi**: Conceptualization, Methodology, Data curation, Writing – review & editing. **Ganesh R Ghimire**: Conceptualization, Methodology, Writing – review & editing, Visualization. **Sanjib Sharma**: Conceptualization, Methodology, Writing – review & editing. **Prithvi Dhwoj Khadka**: Software, Formal Analysis, Writing - review & editing. **Sanghoon Shin**: Methodology, Formal analysis, Writing – review & editing. **Yadu Pokhrel**: Methodology, Formal analysis, Writing – review & editing. **Utsav Bhattarai**: Writing – review & editing. **Rajaram Prajapati**: Writing – review & editing. **Bhesh Raj Thapa**: Writing- review & editing. **Ramesh Kumar Maskey**: Conceptualization, Methodology, Writing – review & editing, Supervision.

**Declaration of Competing Interest**

The authors declare that they have no known competing financial interests or personal relationships that could have appeared to influence the work reported in this paper.

**Data Availability**

Some or all data, models, or codes that support the findings of this study are available from the corresponding author upon reasonable request.

**Acknowledgment**

The authors would like to express sincere gratitude to Mr. Ashish Shrestha, Department of Electricity Development, Nepal for his assistance during the study.